\documentclass[twocolumn,secnumarabic,amssymb, nobibnotes, aps, prl]{revtex4-1}
\usepackage[utf8]{inputenc}
\usepackage{amsmath}
\usepackage{amsfonts}
\usepackage{amssymb}
\usepackage[pdftex]{graphicx}
\usepackage{lmodern}
\usepackage{subfigure}
\usepackage{color}

\begin{document}

\title{Bright 5-85 MeV Compton $\gamma$-ray pulses from GeV laser-plasma electron accelerator and plasma mirror}%

\author{J. M. Shaw, A. C. Bernstein, R. Zgadzaj, A. Hannasch, M. LaBerge, Y. Y. Chang, K. Weichman, J. Welch,  W. Henderson, H.-E. Tsai, N. Fazel, X. Wang, T. Ditmire, M. Donovan, G. Dyer, E. Gaul, J. Gordon, M. Martinez, M. Spinks, T. Toncian,$^{a)}$ C. Wagner and M. C. Downer}%
\affiliation{Department of Physics, University of Texas at Austin, Austin, Texas 78712-1081, USA}
\date{\today}%
\begin{abstract}
We convert a GeV laser-plasma electron accelerator into a compact femtosecond-pulsed $\gamma$-ray source by inserting a $100 \mu$m-thick glass plate $\sim3$ cm after the accelerator exit.  With near-unity reliability, and requiring only crude alignment, this glass plasma mirror retro-reflected spent drive laser pulses (photon energy $\hbar\omega_L = 1.17$ eV) with $>50\%$ efficiency back onto trailing electrons (peak Lorentz factor $1000 < \gamma_e < 4400$), creating an optical undulator that generated $\sim10^8 \gamma$-ray photons with sub-mrad divergence, estimated peak brilliance $\sim10^{21}$ photons/s/mm$^2$/mrad$^2$/$0.1\%$ bandwidth and negligible bremsstrahlung background.  The $\gamma$-ray photon energy $E_\gamma = 4\gamma_e^2 \hbar\omega_L$, inferred from the measured $\gamma_e$ on each shot, peaked from 5 to 85 MeV, spanning a range otherwise available with comparable brilliance only from large-scale GeV-linac-based high-intensity $\gamma$-ray sources.   
\end{abstract}

\maketitle

\noindent  
Atomic nuclei are natural sources of MeV $\gamma$-ray radiation.  However, electron accelerators can generate intense directional beams of MeV photons (known as ``megavoltage x-rays" in some communities) for probing and manipulating the nucleus and for medical, industrial and homeland security applications.  Broadband, unpolarized bremsstrahlung $\gamma$-ray beams, with photon energy $4 < E_\gamma < 25$ MeV, generated in high-$Z$ targets by MeV electrons from small linacs are now standard tools in treatment of deep cancers \cite{Lev12}, sterilization of food and medical equipment, and cargo scanning \cite{Hamm12}.  A wider range of nuclear photonic applications demands polarized, quasi-monochromatic, and/or short-pulsed $\gamma$-ray beams --- \textit{e.g.} 
studies of astrophysical nucleosynthesis mechanisms \cite{Hay06}; non-destructive detection and assay of nuclear materials \cite{Haj09}; isotope-selective transmutation of long-lived fission products \cite{Li09a}; production of medical radioisotopes \cite{Hab11a}; pulsed radiolysis \cite{Tai11}; and efficient generation of ultrashort polarized positron bunches suitable for injection into advanced accelerators \cite{Li09b} --- with some requiring photon energies up to $\sim80$ MeV.  To meet this demand, several GeV-class electron accelerator facilities dedicated to generating intense $\gamma$-ray beams via pulsed laser Compton scatter (LCS) --- \textit{e.g.} the High-Intensity $\gamma$-ray Source (HI$\gamma$S) \cite{Wel09}, NewSUBARU \cite{Ama09} and others \cite{DAn00} --- have been built and operated starting in the 1980s, while new linac-based $\gamma$-ray sources featuring \textit{e.g.} exceptionally narrow bandwidth \cite{Alb11}, high photon flux \cite{Haj08} and ultrashort pulse duration \cite{Tai11} continue to emerge.   These facilities exploit the ability of LCS to map the polarization and spectral-temporal structure of the scattering laser pulse onto the $\gamma$-radiation.  LCS generates the most energetic $\gamma$-rays in the backscatter geometry --- for which $E_\gamma = 4\gamma_e^2\hbar\omega_L$, where $\gamma_e$ is the electron Lorentz factor and $\hbar\omega_L$ the laser photon energy \cite{Esa93}.  Thus to generate 80 MeV $\gamma$-rays via backscatter of a standard Nd$^{3+}$ laser  pulse ($\hbar\omega_L = 1.17$ eV) requires electrons with $\gamma_e = 4.1 \times 10^3$ (energy $E_e = 2.1$ GeV).

Within the past 4 years, compact laser-plasma accelerators (LPAs) \cite{Esa09} have produced 2 to 4 GeV quasi-monoenergetic electron bunches \cite{Wan13,Kim13} within an acceleration distance of centimeters --- thousands of times smaller than conventional GeV linacs.  In these LPAs, an ultrashort drive laser pulse of 0.3 to 0.6 PW peak power traversing cm-length tenuous plasmas blew out positively-charged, light-speed accelerating cavities of $\sim50\mu$m diameter, which captured ambient plasma electrons at their rear and accelerated them in their internal GV/cm electrostatic fields to GeV energy.  The emergence of GeV LPAs raises the intriguing possibility of developing small, easily accessible LCS $\gamma$-sources that span the photon energy range ($1 < E_\gamma < 80$ MeV) of linac-based LCS facilities, complementing their capabilities while more readily providing synchronized electron bunches and $\gamma$-ray pulses of fs duration \cite{Har07}.   Indeed the planned Extreme Light Infrastructure-Nuclear Physics (ELI-NP) facility is based on this possibility \cite{Hab11b}.  


\begin{figure*}[t]
\centering
\includegraphics[trim=.6cm 1.3cm .5cm 1.3cm, width=0.75\textwidth,clip]{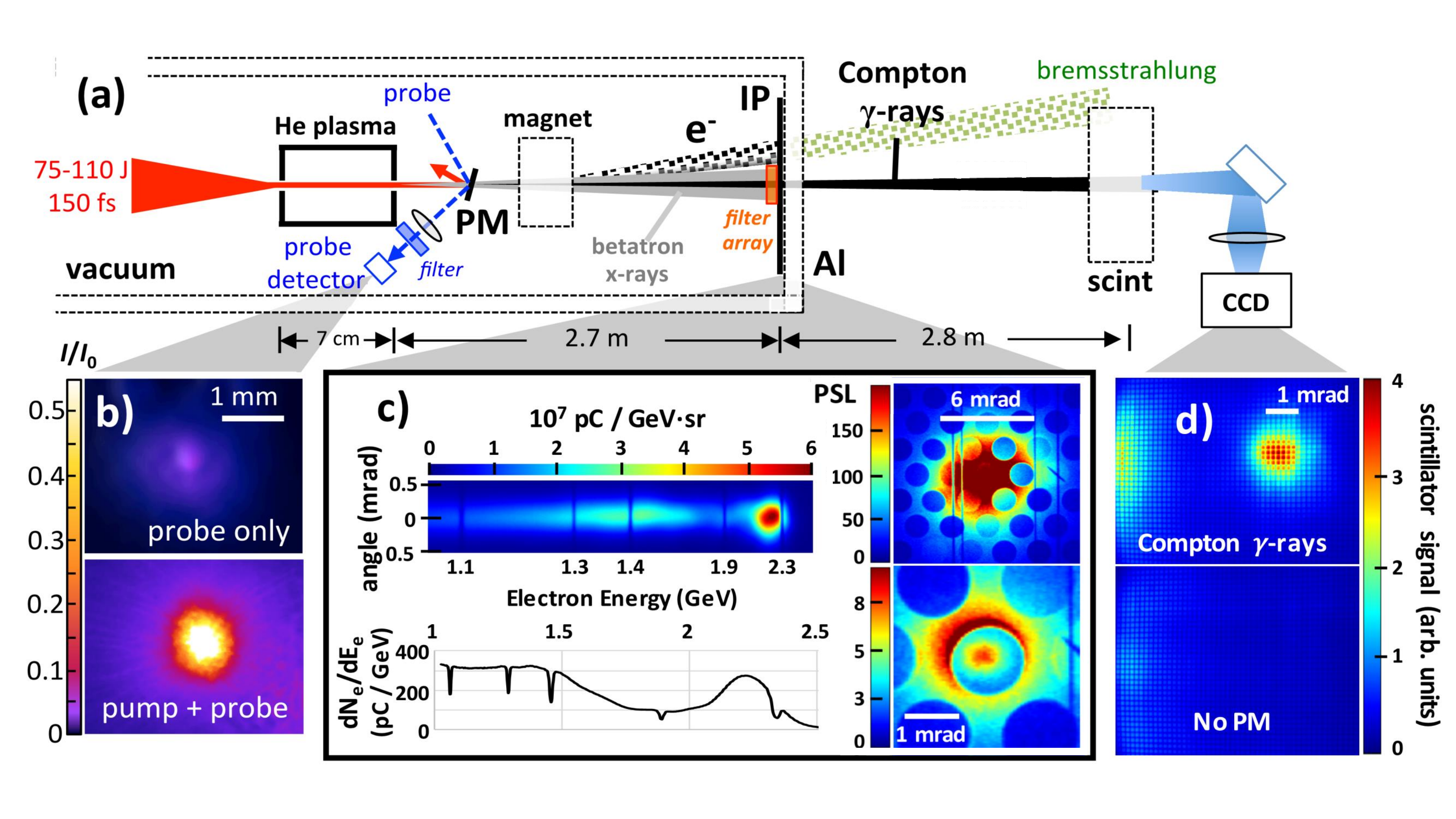}
\caption{Color online. \textbf{(a)} Schematic experimental setup for production and measurement of GeV electrons and Compton $\gamma$-rays:  PM = plasma mirror; scint = pixelated scintillator.  \textbf{(b)} Spatial profile of probe intensity $I$ reflected from PM without (top) and with (bottom) excitation by the transmitted LPA drive pulse, with respect to peak intensity $I_0$ of incident probe. \textbf{(c)} Electron spectrum (left) with peak at 2.2 GeV and corresponding betatron x-ray profile (upper right) recorded on IP.  Tungsten wire fiducials after 1 T magnet imprinted vertical shadows on the electron spectrum to calibrate energy scale \cite{Wan13}.  Dark circles on betatron x-ray profile are thin metal converters (for $\gamma$-rays) or $K$-edge filters (for x-ray energy analysis \cite{Faz16}); secondary particles from $\gamma$-ray conversion produced a bright spot near center of metal disk (lower right) on a separate shot. \textbf{(d)} Scintillator signals with PM in place (top), showing Compton $\gamma$-ray profile, and with no PM (bottom).}
\label{figure1}
\end{figure*}

Previous work based on sub-GeV LPAs has already demonstrated Compton sources up to low-MeV photon energy.  The simplest of these generated broadband \cite{Phu12} or tunable quasi-monochromatic \cite{Tsa15,Dop16,Yu16} Compton backscatter x-rays with measured photon energy up to 2 MeV \cite{Yu16} by inserting a reflective film just after the exit of a $<450$ MeV TW-laser-driven LPA.  The film acted as a plasma mirror (PM) \cite{Gei11} that retro-reflected the intense part of the drive pulse onto trailing accelerated electrons, without alignment difficulty, while the generation of background bremsstrahlung x-rays from LPA electrons was suppressed by using a thin low-Z PM material \cite{Tsa15,Yu16}.  However, this simple technique has not been scaled to GeV LPAs because their PW drive pulses possess stronger pre-pulses (requiring more stringent suppression techniques than TW pulses) that pre-expand the PM surface, degrading its reflectivity and the efficiency and reliability of Compton backscatter.  Instead Compton photons above 2 MeV have been generated from LPAs by the more technically challenging approach of colliding the micrometer-sized LPA electron bunch with a separate pulse created either by splitting off a major fraction of the fully amplified LPA drive pulse (thus limiting the achievable $\gamma_e$) \cite{Che13,Liu14}, or by splitting it from the drive pulse oscillator and separately amplifying it to multi-TW power \cite{Sar14}.  In either case, the backscatter pulse was
focused at the LPA exit through a separate optical path that had to be carefully aligned and compensated for pointing jitter.  Some researchers successfully met these challenges, generating quasi-monochromatic LCS photons up to 9 MeV \cite{Liu14} with a backscattering pulse focused to $a_0 < 1$.  Here $a_0 = eE_L/m_e\omega_Lc$ (where $E_L$ is the backscatter laser electric field, $m_e$ the electron mass) is a dimensionless laser strength parameter that equals $1$ at the threshold of relativistic electron undulation.  Sarri \textit{et al$.$} extended the high-energy tail of the LCS photon spectrum up to 18 MeV by focusing the backscatter pulse to $a_0 > 1$, accessing a nonlinear Compton scatter regime for which the mean $E_\gamma \sim 4\gamma_e^2\hbar\omega_La_0$ \cite{Sar14}.  Since, however, the nonlinear LCS spectrum is inherently composed of multiple harmonics of $E_\gamma = 4\gamma_e^2\hbar\omega_L$, linear backscatter is preferred to produce quasi-monochromatic $\gamma$-rays.  

In this Letter, we report \emph{linear} generation of Compton $\gamma$-rays over the range $5 < E_\gamma < 85$ MeV by inserting a near-retro-reflecting PM (a standard microscope coverslip) $3.3$ cm after the exit of a LPA.  This LPA produced electron bunches of charge $0.05 < q < 0.15$ nC within a quasi-monoenergetic ($.05 < \Delta E_e/E_e < 0.1$) peak that tuned from 0.5 to 2.2 GeV \cite{Wan13}.  By recycling the transmitted drive laser pulse, the PM approach efficiently used available 0.5 to 0.75 PW power \cite{Gau10} for both acceleration and LCS.  Its strength dropped from $a_0 \sim 3$ at the LPA exit \cite{Wan13} to $\sim$0.5 at the PM, where it yielded consistently high ($>50\%$) PM reflectivity and reliable linear LCS.  As a result of a recent upgrade \cite{Gaul16}, our PW drive pulses had sufficient leading edge peak-to-pedestal contrast ($10^5$ at $20$ ps, $>10^8$ at $100$ ps from the peak of each pulse) to overcome pre-expansion of the PM, which would otherwise limit the process \cite{Tsa17}. Although we did not measure $E_\gamma$ directly, we measured the energy distribution of accelerated electrons that produced them with $\pm5\%$ accuracy on every shot using a calibrated magnetic spectrometer \cite{Wan13}.  From the relation $E_\gamma = 4\gamma_e^2\hbar\omega_L$ we inferred peak $E_\gamma$ spanning the entire range currently available from large-scale GeV-linac-based LCS sources \cite{Wel09,Ama09}.  

We carried out experiments at the Texas PW Laser \cite{Gau10,Gaul16}, which provided drive laser (``pump") pulses of 1057 nm center wavelength, 150 fs duration, and energy between 75 and 110 J.  Fig.~1a shows the setup.  A spherical mirror focused the pulses in vacuum at $f/40$ into the entrance aperture of a 7-cm-long gas cell, which we filled uniformly with 6 Torr He gas immediately before each shot.  The pulses fully ionized the gas, creating plasma of electron density $n_e \approx 5\times 10^{17}$ cm$^{-3}$, and generating self-injecting plasma bubbles that accelerated electrons to GeV energy \cite{Wan13}.  The pump pulse transmitted through the accelerator reflected at $7^\circ \pm 2^\circ$ from a PM ($L = 100$ $\mu$m thick fused silica), which we replaced after each shot.  This geometry avoided retro-reflecting the pump into the amplifier chain.  A probe pulse, generated by frequency-doubling a split-off portion of the pump, reflected at $45^\circ$ simultaneously with the pump, and was imaged from the PM to a charge-coupled device (CCD) camera through spectral filters that rejected scattered pump light.

An imaging plate (IP) located at $z = 2.7$ m recorded magnetically-dispersed accelerated electrons, and keV betatron x-rays
 \cite{Rou04}, after they passed through a $50~\mu$m thick aluminum foil (not shown) that deflected any remaining drive pulse into a beam dump.  Energy-dependent electron number $dN_e/dE_e$ and total $q$ were determined from measured photo-stimulated luminescence (PSL) levels scanned from exposed IPs, and quantified using 
a calibration procedure described in Ref.~\cite{Wan13}.  
Compton $\gamma$-rays 
passed through PM, laser deflector, IP, and a 3.3-mm-thick Al back plate of the vacuum chamber (which blocked collinear betatron x-rays) 
before a pixelated CsI(Tl$^+$) scintillator detected them at $z = 5.5$ m.  Calculations of $\gamma$-ray attenuation \cite{NIST} show that secondary particles that the $\gamma$-rays generate in these materials account for $<3\%$ of the scintillator signal.  The $\gamma$-rays alone left no discernible trace on the IP.  However we covered $\sim$ 6 cm$^2$ of the IP with a planar array of forty 4 mm-diameter, 20-200 $\mu$m-thick disks of various metals (Fig. 1c, upper right).  These characterized betatron x-rays \cite{Faz16}, and on some shots, when a $\gamma$-ray pulse passed through one disk, converting a small fraction of its energy to secondary electrons ($e^-$) and positrons ($e^+$) that \emph{did} expose the IP (Fig. 1c, lower right), they allowed us to determine the number $N_\gamma$ of photons in the pulse.  

Figure~1b)-d) presents representative data.  Figure~1b) shows images of the probe reflected from the PM without (top) and with (bottom) simultaneous irradiation by the spent LPA drive pulse.  Within the 1-mm-diameter pump-irradiated region, PM reflectivity increased ten-fold.
Based on the size of the imaged high-reflectivity region, we conclude that the transmitted pump diverged at cone angle $\theta_L \sim 30$ mrad (FWHM), and had intensity $I \alt 4 \times 10^{17}$ W/cm$^2$ ($a_0 \lesssim 0.5$) incident on the PM.

Figure 1c) shows a magnetically dispersed electron spectrum (left) with a peak at $E_e=2.2$ GeV and angular divergence $\theta_e \approx0.9$ mrad (FWHM), and betatron x-rays (right) of angular divergence $\theta_x \approx6$ mrad (FWHM), recorded on the IP. Figure 1d) shows the $\sim$5 mm diameter spatial profile of $\gamma$-rays recorded on the $5 \times 5$ cm scintillator (top), corresponding to angular divergence $\theta_\gamma \approx1$ mrad.  With the PM in place, we observed a signal of similar brightness and $\theta_\gamma$ from every shot that produced quasi-monoenergetic GeV electrons.  With no PM, we observed no such signal (bottom), even for shots that generated copious GeV electrons.  The diffuse signal at the left edges of the panels in d) is forward bremsstrahlung generated by the least magnetically-deflected (highest $E_e$) electrons in the vacuum chamber back plate.  We observe this signal regardless of the presence of the PM.   


\begin{figure}[htb]
\centering
\includegraphics[trim=0.5cm 0.5cm 0.5cm 0.3cm, width=0.35\textwidth]{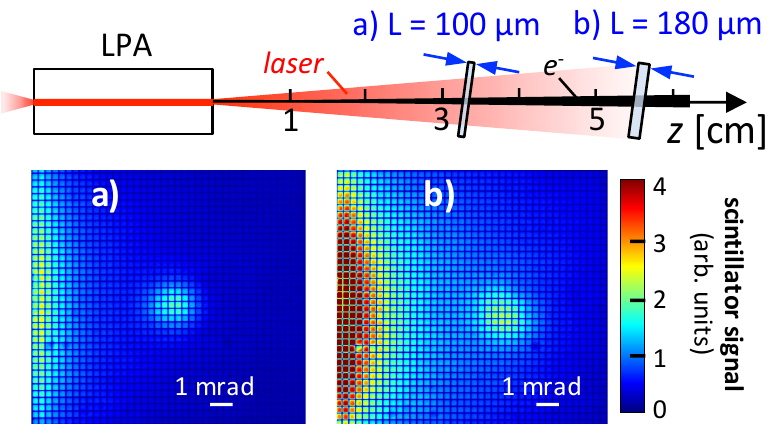}
\caption{Color online. Scaling of scintillator signal with position $z$ and thickness $L$ of PM:  (a) $z = 3.3$ cm, $L = 100 \mu$m; (b) $z = 5.5$ cm, $L = 180$ $\mu$m.  Nearly identical laser pulses drove both shots; both yielded electrons with energy peaked at 0.92 GeV and corresponding charge (a) 50 or (b) 125 pC.}
\label{figure2}
\end{figure}

Observed $\gamma$-ray signals that depend on the PM could be generated from GeV electrons either by forward bremsstrahlung radiation \emph{within} the PM or by LCS \emph{in front of} the PM.  To distinguish these possibilities, we observed how the scintillator signal depended on $L$, PM material, and intensity $I_{\rm R}(z)$ reflected from the PM.  Bremsstrahlung is proportional to $L$ and increases for PM materials of higher $Z$, but does not depend on $I_{\rm R}(z)$.   LCS, on the other hand, does not depend on PM thickness or material, but is proportional to $I_{\rm R}(z)$, which we varied by adjusting the distance $z$ over which the spent drive pulse diverged from accelerator to PM. This intensity in turn determined the PM reflectivity \cite{Tsa17}.  

As an example, Fig.~2 compares scintillator signals from two shots driven by nearly identical laser pulses that yielded electron bunches spectrally peaked at $E_e = 920\pm20$ MeV with total charge $q$ = 50 (a) or 125 pC (b).  The integrated signals in Fig. 2a) and 2b), which were generated with glass PMs of $L = 100$ (a) or $180$ $\mu$m (b) located at $z_a = 3.3$ (a) or $z_b = 5.5$ cm (b), respectively, as illustrated at the top of Fig.~2, have the ratio $S_b/S_a = 1.3$.  Normalized to $q$, with which both bremsstrahlung and LCS scale linearly, the ratio becomes $[S_b/S_a]_n = S_b/S_a \times q_a/q_b = 1.3 \times 0.4 = 0.52$. Signals dominated by bremsstrahlung would have yielded $[S_b/S_a]_n = 1.8$ in view of the $1.8\times$ thicker PM in case (b).  On the other hand, signals originating mostly from LCS should yield $[S_b/S_a]_n = I_{\rm R}(z_b)/I_{\rm R}(z_a)$.  The estimated squared field strength $a_0^2(z) \propto z^{-2}$ \emph{incident} on each PM was $a_0^2(z_a) \approx 0.25$ and $a_0^2(z_b) \approx 0.09$, which yield slightly different PM reflectivities $R_a \approx 0.7$ and $R_b \approx 0.9$ \cite{Tsa17}.  Thus we expect $I_{\rm R}(z_b)/I_{\rm R}(z_a) = (0.9/5.5^2) / (0.7/3.3^2) \approx 0.46$, in good agreement with the observed $[S_b/S_a]_n$. 
For comparison, the bremsstrahlung signal at the left-hand edge of panel (b) is $2.4\times$ stronger than its counterpart in panel (a), a consequence of the $2.5\times$ higher $q$.

\begin{figure}[htb]
\centering
\includegraphics[trim=0 0 0 0, width=0.4\textwidth,clip]{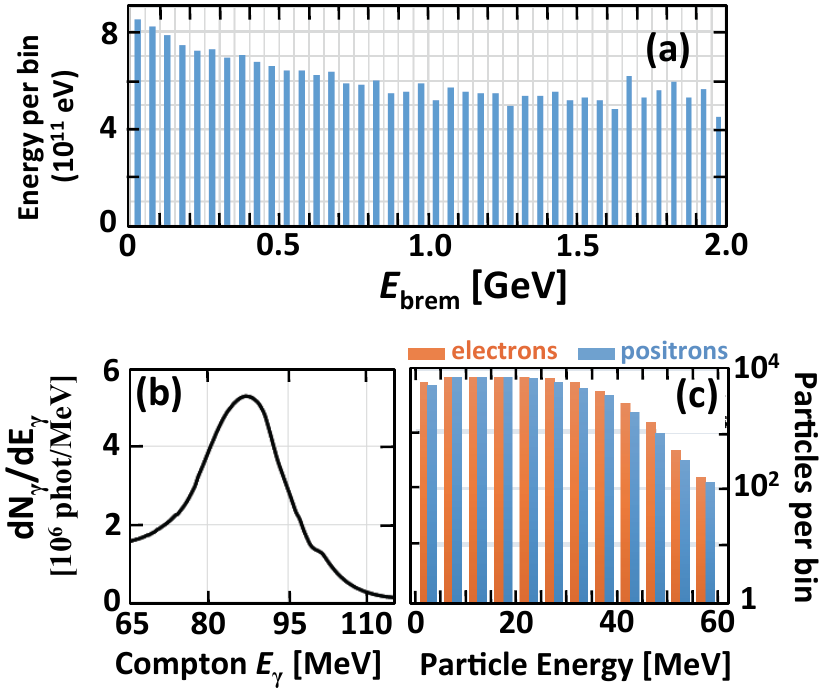}
\caption{Color online. Calculated properties of $\gamma$-rays from PM.  \textbf{(a)}  GEANT4 calculation of bremsstrahlung spectrum produced by 2.2 GeV electrons in 100$\mu$m glass PM.   \textbf{(b)}  $dN_\gamma/dE_\gamma$ of 85 MeV Compton $\gamma$-rays, calculated from $dN_e/dE_e$ in Fig.~1c (left) and Eq.~(1).  \textbf{(c)} GEANT4 calculation of $e^-$$e^+$ energy distributions produced by 40 MeV $\gamma$-rays in converter that led to IP exposure in Fig.~1c (lower right).  }
\label{figure3}
\end{figure}

Analysis of other shots, and calculations, supported the conclusion that LCS dominated for our conditions.  For example, a separate series of $\sim 20$ shots using low-$Z$ (plastic) PMs with $L$ varying from 12.5 to 125 $\mu$m yielded normalized scintillator signals similar to those in Fig.~\ref{figure2}, with no discernible $L$-dependence.   As a second example, the calculated bremsstrahlung energy loss of the 50 pC of electrons within the $E_e \approx 2.2$ GeV ($\Delta E_e^{\rm (FWHM)} = 0.25$ GeV) peak in Fig.~1c traveling through $L = 100$ $\mu$m fused silica (density $\rho_{\rm SiO_2} = 2.5$ g/cm$^3$) is $E_{brem} \approx \alpha_B \rho_{\rm SiO_2}qL \approx 6.2 \times 10^{14}$ eV, where $\alpha_B \approx80$ MeV-cm$^2$/g is the radiative stopping power of 2.2 GeV electrons in silica \cite{NIST}.  The simulation toolkit GEANT4 \cite{All06} yielded similar $E_{brem} \approx 7.4 \times 10^{14}$ eV, along with the spectrum in Fig.~\ref{figure3}a.  For comparison, the spectral density of LCS within the observed solid emission angle $\Delta\Omega = 0.92$ $\mu$sr (Fig.~1d, top) is \cite{Sch06}
\begin{equation} \label{eq:1}
\frac{dN_\gamma}{dE_\gamma} \approx \alpha_f \frac{\gamma_e a_0^2}{8\hbar\omega_L} \Delta\Omega\left[\frac{dN_e}{dE_e} \right],
\end{equation} 
where $\alpha_f$ is the fine structure constant, and (for data in Fig.~1c,d) $\gamma_e = 4400$, $a_0 \approx0.3$ retro-reflected from the PM.  
Fig.~\ref{figure3}b shows the resulting $dN_\gamma/dE_\gamma$ peaked at $E_\gamma = 85$ MeV with $\Delta E_\gamma^{\rm (FWHM)} = 18$ MeV.  The FWHM contains $N_\gamma = 7.6 \times 10^{7}$ photons of total energy $\sum (N_\gamma \times E_\gamma) = 6.8 \times 10^{15}$ eV, about $11\times$ the estimated bremsstrahlung yield in the PM.  This supports the conclusion that most observed $\gamma$-rays originate from LCS.  The entire $dN_e/dE_e$ in Fig.~1c generated a total $N_\gamma \approx2 \times 10^8$ photons from $20 < E_\gamma < 100$ MeV.


We verified the accuracy of Eq.~(\ref{eq:1}) for estimating $dN_\gamma/dE_\gamma$ by analyzing the secondary $e^-$$e^+$ yield of a 40 MeV Compton $\gamma$-ray pulse that produced the IP exposure in Fig.~1c (lower right) upon passing through a Ag (75$\mu$m)/Cu (34$\mu$m) film pair.  We input the $dN_\gamma/dE_\gamma$ from this shot (not shown), calculated from the measured $dN_e/dE_e$ via Eq.~(\ref{eq:1}), into GEANT4, approximating it as  using a gaussian peaked at $E_\gamma = 40$ MeV with $\Delta E_\gamma^{\rm (FWHM)} = 20$ MeV.  Upon passing through the converters, the simulated $\gamma$-ray pulse (total $N_\gamma \sim 10^7$) generated $e^-$ and $e^+$ energy distributions shown in Fig.~\ref{figure3}c with total particle number $(N_e^- + N_e^+) \approx1.28 \times 10^5$.  We compared this with the total particle number $N_{IP}$ obtained directly from the measured PSL within the FWHM of the exposed spot in Fig.~1c, using published MS IP sensitivity $\approx20\pm5$ mPSL/e \cite{Rab16}.  The result --- $N_{IP} \approx1.1 \pm 0.2 \times 10^5$ ---  agreed well with the simulated value.    


\begin{figure}[htb]
\centering
\includegraphics[trim=0 0 0 0,width=0.45\textwidth,clip]{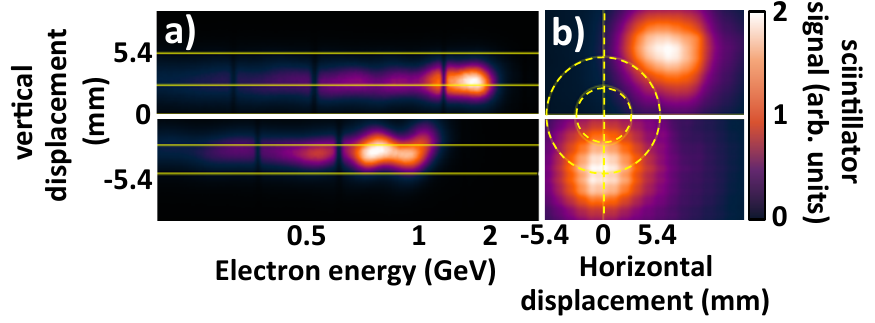}
\caption{Color online.  Shot-to-shot pointing and divergence fluctuations. \textbf{(a)} Electron spectra and \textbf{(b)} $\gamma$ profiles, showing equal and opposite vertical angle displacements, and different horizontal displacements, on each of two shots. }
\label{figure4}
\end{figure}

Shot-to-shot fluctuations in the pointing and angular divergence of the Compton $\gamma$-ray beam closely tracked corresponding fluctuations of the GeV electron beam, and thus provided valuable e-beam diagnostics. As an example, Fig.~\ref{figure4} shows matching $\pm 1$ mrad \emph{vertical} angular displacements of (a) electrons ($\pm 2.7$ mm at $z = 2.7$ m) and (b) $\gamma$-rays ($\pm 5.4$ mm at $z = 5.4 m$) for two shots with respect to a common horizontal alignment plane (solid horizontal white line).  The simultaneous observation of \emph{horizontal} $\gamma$-deflections differing by $\sim 6$ mrad between the two shots (see Fig.~\ref{figure4}b) demonstrates different horizontal $e^-$ deflections, a fact not easily discerned from the horizontally-dispersed electron spectra themselves.  Thus the $\gamma$-centroid diagnoses electron launch angle from the LPA, an essential parameter in calibrating the magnetic spectrometer, more accurately and directly than 2-screen electron detection methods \cite{Cla10,Wan13}.  In addition, we plotted the vertical divergence (half-cone) angles $\theta_e$ ($\theta_\gamma$) of electron ($\gamma$)-beams for many shots and found that they fit well ($R^2 = 0.94$) to a linear function $\theta_\gamma = 1.26\theta_e + 0.4$ mrad.  The non-zero intercept matches the expected divergence angle $\sim 1/\gamma_e$ for Compton backscatter from a \emph{single} $\gamma_e = 2500$ electron.  For $\theta_e > 0$, $\theta_\gamma$ is a convolution of $\theta_e$ and $1/\gamma_e$.  The strong correlation between $\theta_\gamma$ and $\theta_e$ suggests that quantitative analysis of the $\gamma$-ray profiles may help to diagnose transverse emittance of GeV electron bunches on each shot in future studies \cite{Cho06}.  


\begin{figure}[htb]
\centering
\includegraphics[trim=0.2cm 0.2cm 0.0cm 0.8cm,width=0.45\textwidth,clip]{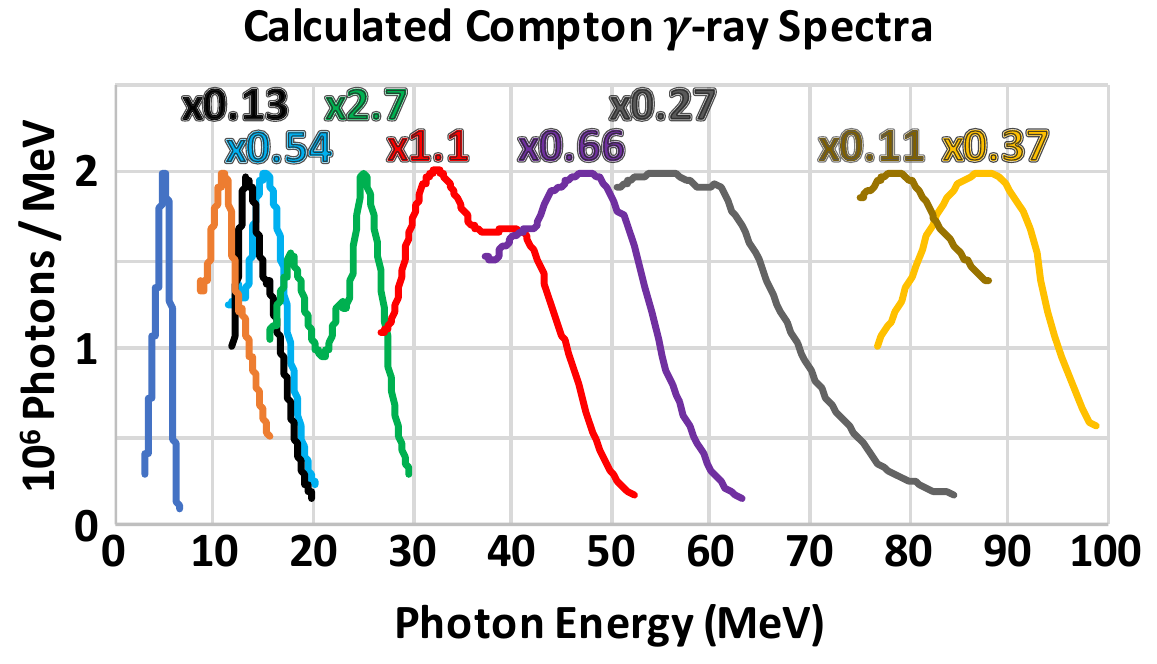}
\caption{Color online.  Quasi-monochromatic Compton $\gamma$-ray spectra generated as $E_e$ tuned from 0.5 to 2.2 GeV.  Spectra are labeled with multipliers that normalize true peak heights to the height of the two lowest energy curves. }
\label{figure5}
\end{figure} 

Fig.~5 displays ten quasi-monochromatic Compton $\gamma$-ray spectra peaking from 5 to 85 MeV, determined from measured $dN_e/dE_e$ using Eq.~1.   Over the tuning range, $\theta_\gamma$ (half-cone) averaged $1 \pm 0.3$ mrad and $\Delta E_\gamma^{(FWHM)}$ of the spectral peaks averaged $15 \pm 11$ MeV, where variances denote standard deviation.  Drive pulse energy was the primary control variable in tuning $E_e$ and $E_\gamma$, although shot-to-shot fluctuations in pulse duration, focal profile, and wavefront also played a role.  For all but two of the peaks, $N_\gamma$ within the FWHM ($0.1\%$ bandwidth) fell in the range $0.3$ to $14 \times 10^7$ ($0.2$ to $8.8 \times 10^5$). For the two exceptions, $N_\gamma$ was \emph{greater}:  45 and 173 $\times 10^7$ (36.5 and 38.6 $\times 10^5$) within the FWHM ($0.1\%$ bandwidth).  Combining these numbers with estimated $\gamma$-ray pulse duration $\tau_\gamma \approx80$ fs (\textit{i.e.} half a plasma period for $n_e \approx 5\times 10^{17}$) and estimated source-size radius of 20 $\mu$m at $z = 3.3$ cm, we estimate peak-brilliance ranging from $(0.1-4.6)\times 10^{21}$  photons/s/mm$^2$/mrad$^2$ within $0.1\%$ bandwidth.  The highest number originates from the most energetic spectral peak in Fig.~5, whose $dN_\gamma/dE_\gamma$ is presented in Fig.~4b.  

Currently energy spread of our LPA electrons ($\Delta E_e^{(FWHM)} \sim 100$ MeV) limits $\Delta E_\gamma^{(FWHM)}$.  However, LPAs have produced $\Delta E_e^{(FWHM)} < 3$ MeV at $E_e = 180$ MeV \cite{Rec09} and $< 0.2$ MeV at $E_e \alt 1$ MeV \cite{Ged08}, using specialized injection methods. Simulations show such widths can be preserved to GeV energy \cite{Ged08}.   Moreover, $a_0^2$ of the backscatter pulse (and thus $N_\gamma$) can be increased approximately ten-fold, while LCS remains linear and PM reflectance high \cite{Tsa17}, by moving the PM closer to the LPA exit.  These combined improvements could decrease $E_\gamma$ (FWHM) to $\sim 0.1$ MeV, opening up nuclear spectroscopy applications, while increasing peak brilliance to $>10^{23}$ photons/s/mm$^2$/mrad$^2$/$0.1\%$ bandwidth.

In summary, we converted a compact 0.5-2.2 GeV laser-plasma electron accelerator into a bright fs-pulsed, quasi-monochromatic Compton $\gamma$-ray source with peak photon energies tunable from 5 to 85 MeV by inserting a low-$Z$ plasma mirror near the accelerator exit.  We foresee such sources eventually complementing large linac-based LCS sources in nuclear photonic research and applications.  

DOE grants DE-SC0011617 and DE-SC0012444, AFOSR grant FA9550-14-1-0045 and NNSA Cooperative Agreement DE-NA0002008 supported this work.  KW was supported by Computational Sciences Graduate Fellowship under DOE grant DE-FG02-97ER25308. 

${^a)}$ Present address:  Institute for Radiation Physics, Helmholtz-Zentrum Dresden-Rossendorf, Germany.

\end{document}